\documentclass[twocolumn,prb,aps,floatfix,epsf,psfig,superscriptaddress,amssymb]{revtex4}
\usepackage{epsfig}
\usepackage{graphics}
\usepackage{amsmath}
\usepackage{txfonts}

\begin{document}

\title{Zero-energy states of graphene triangular quantum dots in a magnetic field}

\author{A.~D.~G\"u\c{c}l\"u}
\affiliation{Department of Physics, Izmir Institute of Technology, IZTECH,
  TR35430, Izmir, Turkey}
\author{P.~Potasz}
\affiliation{Institute of Physics, Wroclaw University of Technology, Wroclaw, Poland}

\author{P.~Hawrylak}
\affiliation{Institute for Microstructural Sciences, National Research Council of Canada
, Ottawa, Canada}
\date{\today}

\begin{abstract}
We present a tight-binding theory of triangular graphene quantum dots
(TGQD) with zigzag edge and broken sublattice symmetry in external
magnetic field. The lateral size quantization opens an energy gap and
broken sublattice symmetry results in  a shell of degenerate states at
the Fermi level. We derive a semi-analytical form for zero-energy
states in  a magnetic field and show that the shell remains degenerate
in a magnetic field, in analogy to the 0th Landau level of bulk
graphene. The magnetic field closes the energy gap  and leads to the
crossing of valence and conduction states with the zero-energy states,
modulating the degeneracy of the shell. The closing of the gap with
increasing magnetic field is present in all graphene quantum dot
structures investigated irrespective of shape and edge termination.
\end{abstract}

\maketitle
\section{INTRODUCTION}
Graphene currently attracts considerable attention due to remarkable
electronic and mechanical
properties.\cite{Wallace+47,Novoselov+Geim+04,Novoselov+Geim+05,Zhang+Tan+05,Son+PRL+06,Potemski+deHeer+06,Geim+Novoselov+07,Rycerz+Tworzydlo+07,Xia+Mueller+09,Mueller+Xia+10,Neto+Guinea+09}
When graphene is reduced to graphene nanostructures, new effects
related to size-quantization and edges
appear.\cite{Neto+Guinea+09,Abergel+10,Rozhkov+11} Considerable
experimental effort has been made aiming at producing graphene
nanostructures with desired shape and
edges.\cite{Li+08,Ponomarenko+08,Ci+08,You+08,Schnez+09,Ritter+09,Jia+09,Campos+09,Neubeck+10,Biro+10,CruzSilva+10,Yang+10,Krauss+10,Zhi+08,Treier+10,Mueller+10,Morita+11,Lu+11,Chen+12}
Among graphene nanostructures, nanoribbons and quantum dots are of
particular interest. In graphene quantum dots, a size-dependent energy
gap opens,\cite{Yamamoto+06,Zhang+08,Guclu+10} and its magnitude is
determined by shape and edge termination. In graphene quantum dots
with zigzag type edges, edge states with energy in the vicinity of the
Fermi energy appear.
\cite{Son+PRL+06,Yamamoto+06,NFD+96,Fujita+96,Son+06,Ezawa+06,Ezawa+07,FRP+07,AHM+08,Wang+Yazyev+09,Wimmer+10,Potasz+10,Voznyy+11}
These edge states have significant effects on low-energy electronic
properties such as a decrease of the energy gap compared to structures
with armchair termination or, when combined with broken sublattice
symmetry, a creation of the degenerate shell of zero-energy states in
the middle of the energy gap.
\cite{Yamamoto+06,Ezawa+07,FRP+07,AHM+08,Wang+Yazyev+09,Guclu+09,Wimmer+10,Potasz+10,Voznyy+11,Ezawa10,Wang+Meng+08}
It was shown that the degenerate shell survives when various types of disorder are present in the system.\cite{Wimmer+10,Potasz+10,Voznyy+11,Ezawa10}

The influence of an external magnetic field on the electronic
properties of the graphene quantum dots was also
studied. \cite{Yazyev10,Chen+07,Schnez+08,Recher+07,Abergel+08,Wurm+08,Guttinger+09,Schnez+09,Libisch+10,Zhang+08,Grujic+11,Zarenia+11,Bahamon+Pereira+09,Romanovsky+11,Romanovsky+12,Potasz+09}
The magnetic field plays the role of a tunable external parameter
allowing to change electronic properties in a controllable
way. Graphene quantum dots and rings with circular, square, hexagonal,
triangular, and rhombus-shaped shapes with zigzag and armchair edges
were
investigated.\cite{Recher+07,Abergel+08,Wurm+08,Schnez+09,Bahamon+Pereira+09,Libisch+10,Grujic+11,Zarenia+11,Romanovsky+11}
Triangular graphene quantum dots with reconstructed edges, consisting
of a succession of pentagons and heptagons, were also
considered.\cite{Romanovsky+12} The comparison between tight-binding
and continuum model, the Dirac-Weyl equation, was  analyzed for
graphene quantum dots with different type of edges: zigzag, armchair,
and infinite-mass boundary
conditions.\cite{Grujic+11,Zarenia+11,Romanovsky+11} For a circular
dot, good qualitative agreement between experiment and analytical
model with infinite-mass boundary conditions was
obtained.\cite{Schnez+08,Schnez+09} Magneto-optical properties were
also theoretically investigated.\cite{Zhang+08,Grujic+11} The
absorption spectra differ for hexagonal structures with armchair and
zigzag edges due to different level structures and the oscillator
strengths. A fast reduction of the energy gap with increasing magnetic
field in zigzag hexagon in comparison with zigzag triangle was
noted.\cite{Grujic+11,Zarenia+11}

In this work, we present  a tight-binding theory of triangular
graphene quantum dots(TGQD) with zigzag edge and broken sublattice
symmetry in external magnetic field. The lateral size quantization
opens an energy gap and broken sublattice symmetry results in  a shell
of degenerate states at the Fermi level. Building on our previous
work\cite{Potasz+10} we derive here a semi-analytical form for
zero-energy states in  a magnetic field and show that the shell
remains degenerate at all magnetic fields perpendicular to the plane
of the TGQD, in analogy to the 0th Landau level of bulk
graphene. However, we find that the magnetic field closes the energy
gap  and leads to the crossing of valence and conduction states with
the zero energy states, modulating the degeneracy of the shell. The
closing of the gap with increasing magnetic field is present in all
graphene quantum dot structures investigated irrespective of shape and
edge termination.

The paper is organized as follows. In Sec. II, we present a brief
outline of the tight-binding model with an incorporation of a
perpendicular magnetic field. The analysis of the evolution of the
energy spectra of TGQD, a derivation of the analytical form for
eigenfunctions corresponding to zero-energy states, and a prediction
of crossings of valence and conduction states with the zero energy
Fermi level $E=0$ are included in Sec. III. In Sec. IV the energy gap
in a magnetic field for GQDs with different shapes and edge
termination is considered. The conclusions are presented in Sec. V.

\section{MODEL}
We describe graphene quantum dots using the nearest-neighbor
tight-binding model which has been successfully used to describe
graphene  \cite{Wallace+47} and applied to other graphene materials
such as nanotubes, nanoribbons and quantum dots
\cite{NFD+96,Ezawa+06,Yamamoto+06,Ezawa+07,FRP+07,AHM+08,Potasz+10,Saito+98}. A
perpendicular magnetic field can be incorporated by using Peierls
substitution \cite{Peierls+33}. The Hamiltonian reads,
\begin{eqnarray}
  H_{TB}= t\sum_{\left\langle i,j\right\rangle,\sigma}
  \left[ e^{i\varphiup_{ij}} a^\dagger_{i\sigma}b_{j\sigma}
       +e^{-i\varphiup_{ij}} b^\dagger_{j\sigma}a_{i\sigma}\right],
\label{HTB}
\end{eqnarray}
where $t$ is hopping integral, $a^\dagger_{i\sigma}$($b^\dagger_{i\sigma}$)
and $a_{i\sigma}$($b_{i\sigma}$) are creation and annihilation operators on a
site $i$ corresponding to sublattice A(B) of bipartite honeycomb lattice,
$\left\langle i,j\right\rangle$ indicate summation over nearest-neighbors, and
$\sigma$ is spin index. Hopping integral between nearest neighbors is $t=-2.8$
eV \cite{Neto+Guinea+09}. Under symmetric gauge, a vector potential
${\bf{A}}=B_z/2(-y,x,0)$, and
\begin{eqnarray}
\varphiup_{ij}=2\pi\frac{e}{hc}\int_{r_i}^{r_j}{\bf A}d{\bf l}
=2\pi\frac{B_z}{2\phi_0}\left(x_iy_j-x_jy_i\right)
\label{phaseij}
\end{eqnarray}
corresponds to a phase accumulated by electron going from site $i$ to $j$,
which is equal to a magnetic flux going through area
$S=\frac{x_iy_j-x_jy_i}{2}$ spanned by vectors ${\bf r}_i$ and ${\bf r}_j$, and
$\phi_0=\frac{hc}{e}$ is magnetic flux quantum. The evolution of the energy
spectrum in a magnetic field will be shown in units of the magnetic flux
threading one benzene ring, $\phi/\phi_0=B_zS_0/\phi_0$, where
$S_0=3\sqrt{3}a_0^2/2$ is benzene ring area with $a_0=1.42$ $\AA$.

\section{ZIGZAG TRIANGULAR QUANTUM DOT IN A MAGNETIC FIELD}

\subsection{The evolution of the energy spectrum}
\begin{figure}
\epsfig{file=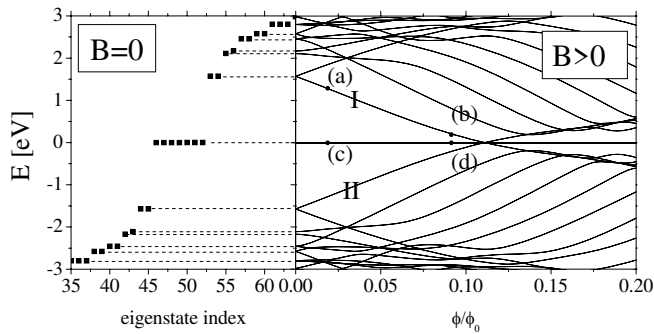,width=3.4in}
\caption{Left: Energy spectrum of triangular graphene quantum dot with
  $N=97$ atoms and $N_{ed}=7$ degenerate zero-energy states in the
  absence of a magnetic field. Right: the evolution of the spectrum
  from the left in a magnetic flux. The degenerate zero-energy shell
  is immune to the magnetic field. The states from the conduction and
  valence bands, labeled by I and II, respectively, crosses for
  $\phi/\phi_0\simeq 0.11$, closing the energy gap.}
\label{fig:Fig1}
\end{figure}

We focus here on the effect of the magnetic field on the electronic
properties of TGQDs, quantum dots with broken sublattice symmetry.  We
illustrate the energy spectrum and its evolution with increasing
magnetic field on a TGQD with $N=97$ carbon atoms.
Fig. \ref{fig:Fig1} shows the energy spectrum  and it's evolution in
the magnetic field obtained by numerical diagonalization of the
Hamiltonian, Eq. (\ref{HTB}).

\begin{figure}
\epsfig{file=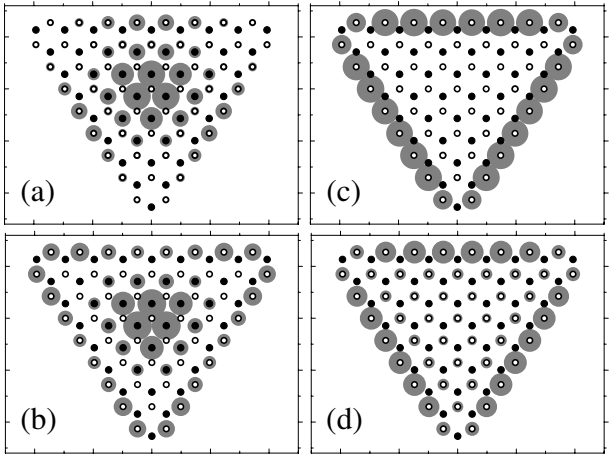,width=3.4in}
\caption{(Color online) The evolution of electronic densities in a
  magnetic field of the lowest state from the conduction band labeled
  by $I$, (a) and (b), and the degenerate shell of $N_{deg}=7$ energy
  levels (the sum of electronic densities of all $N_{deg}=7$
  degenerate states), (c) and (d). The radius of gray circles is
  proportional to the electronic probability density on a given
  site. (a) The state $I$ for $\phi/\phi_0\simeq 0.01$ is mostly
  localized in the center of the dot and with increasing magnetic
  field starts to occupy also edges, shown for $\phi/\phi_0\simeq
  0.08$ in (b). (c) The degenerate states for $\phi/\phi_0\simeq 0.01$
  are strongly localized on edges and for $\phi/\phi_0\simeq 0.08$
  move slightly to the center of the triangle, shown in (d).}
\label{fig:Fig2}
\end{figure}

At $B=0$ there are $N_{deg}=7$  degenerate states at zero energy or
Fermi level. The number of states is equal to the difference between
the number of $A$ and $B$ atoms \cite{Potasz+10}. The states belonging
to the degenerate shell are primarily localized at the edge of the
triangle and are entirely localized on one sublattice, say A, as shown
in Fig. \ref{fig:Fig2}(c).

The evolution of the energy spectrum as a function of the magnetic
field is shown on the right hand side of Fig. \ref{fig:Fig1}. The
spectrum is symmetric with respect to $E=0$ due to electron-hole
symmetry. This symmetry is broken when hoppings to the second nearest
neighbors in Hamiltonian,Eq. (\ref{HTB}), are included. The highest
valence state and the lowest conduction state with $E=\pm 1.57$ eV,
which in the absence of the magnetic field are each doubly degenerate,
split in the presence of a magnetic field. The state labeled by II
from the valence band increases and the state labeled by I from the
conduction band decreases its energy with increasing magnetic field,
closing the energy gap. Around $\phi/\phi_0\simeq 0.11$ these states
reach Fermi level at $E=0$.

The explanation of why the energy gap closes in a magnetic field can
be found by considering Dirac Fermions in bulk graphene.
\cite{Novoselov+Geim+05,Zhang+Tan+05}. We focus on one of two Fermi
points, say $K$ point. Following Refs. \onlinecite{mcclure+56,Toke+06}
the energy spectrum of Dirac Hamiltonian in the presence of magnetic
field is given by
\begin{eqnarray}
E_n=\pm\sqrt{2\hbar v_FeB_z|n|/c},
\label{LLDirac}
\end{eqnarray}
where $v_F$ is Fermi velocity, $c$ speed of light, and $n$ Landau
level index. The $\pm$ sign corresponds to electron (hole) Landau
levels. A unique property of the energy spectrum is the existence of
the $n=0$   Landau level  with energy $E=0$,  constant for all
magnetic fields. When the magnetic field is applied to graphene
quantum dots, discrete energy levels evolve into the degenerate Landau
Levels for Dirac Fermions. Thus, some  levels have to evolve into the
$0$-th Landau level, closing the energy gap as shown in
Fig. \ref{fig:Fig1}. Another feature of the $0$-th Landau level is
that the wavefunctions are localized on only one sublattice, similar
to the zero-energy states in TGQD\cite{Potasz+10}.


 We note in Fig. \ref{fig:Fig1} that the zero-energy degenerate shell
 is immune to the magnetic field as is the $n=0$ Landau level. This is
 certainly different from electronic states in semiconductor quantum
 dots, where $\sim B^2$ dependence is observed.  \cite{raymond+04}

These comments are now illustrated by examining wavefunctions of a
TGQD in a magnetic field.  We investigate the evolution of the
probability density of the wavefunction corresponding to state I,
bottom of the conduction band, from Fig. \ref{fig:Fig1}, and the total
probability density of the zero-energy degenerate shell in a magnetic
field. For state I, probability densities at low and high magnetic
field values are shown in Fig. \ref{fig:Fig2}(a) and (b),
respectively. We note that due to the electron-hole symmetry, an
identical evolution for the state II from Fig. \ref{fig:Fig1} (not
shown here) occurs. Eigenfunctions of states with energy $-|E|$ and
$+|E|$ differ only by a sign of a coefficient on sublattice $B$
indicated by filled circles in Fig. \ref{fig:Fig2}, giving identical
electronic densities. For $\phi/\phi_0\simeq 0.01$,
Fig. \ref{fig:Fig2}(a), the state I is mostly localized at the center
of the dot. With increasing magnetic field, it starts to occupy the
edge sites, shown for $\phi/\phi_0\simeq 0.08$. We note that for
arbitrary magnetic field this state is equally shared over two
sublattices, {\it i.e}, has $50\%$ sublattice content. In
Fig. \ref{fig:Fig2}(c) and (d) the evolution of the total electronic
density of the degenerate zero-energy shell is shown. The electronic
density of the degenerate shell is obtained by summing over all
$N_{deg}=7$ states. Initially, degenerate states are strongly
localized on edges, shown in Fig. \ref{fig:Fig2}(c) for
$\phi/\phi_0\simeq 0.01$. When the magnetic field increases, these
states move slightly towards the center of the triangle, shown in
Fig. \ref{fig:Fig2}(d) for $\phi/\phi_0\simeq 0.08$. We note that even
in the presence of an external magnetic field states from the
degenerate shell are still localized on only one type of atoms,
sublattice A, indicated by open circles in Fig. \ref{fig:Fig2}.

\subsection{Analytical solution for zero-energy states}
Fig. \ref{fig:Fig1} shows that  numerical diagonalization of the
tb-Hamiltonian gives the zero-energy states  immune to external
magnetic field. We will now prove this  analytically. Our first goal
is to show the existence of and find an expression for zero-energy
eigenstates in the presence of a magnetic field. The zero energy
states, if they exist, must be solutions of the singular eigenvalue
problem,
\begin{eqnarray}
H_{TB}\Psi=0,
\label{singul}
\end{eqnarray}
where the Hamiltonian $H_{TB}$ is given by Eq. (\ref{HTB}). There is no coupling
between two sublattices and the solution can be written separately for
$A$-type and $B$-type of atoms. We first focus on sublattice $A$ with an
eigenfunction given by
\begin{eqnarray}
|\Psi^{A}\rangle=\sum_j C_{j}a^\dagger_{j}|0\rangle=\sum_j C_{j}|\phi_{j}^{A}\rangle,
\label{psiA}
\end{eqnarray}
where $C_{j}$ are expansion coefficients of eigenstates written in a basis of
$p_z$ orbitals $\phi_{j}^{A}$ localized on $A$-type site $j$ for either spin state omitted in what follows.

\begin{figure}
\epsfig{file=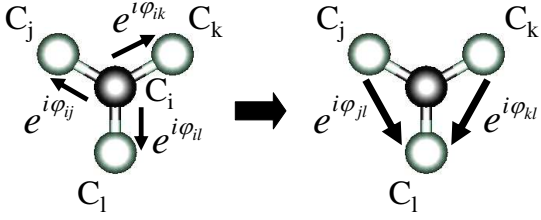,width=3.0in}
\caption{(Color online) Left: phase changes going from $B$-type site
  $i$ to three nearest neighbors $A$-type sites $j$, $k$, and $l$. The
  sum of three $A$-type coefficients multiplied by corresponding phase
  changes must vanish for zero-energy states. Right: phase changes
  going from $A$-type site $k$ to $j$, and $l$ to $j$. A coefficient
  from the bottom, $C_{l}$, can be expressed as a sum of coefficients
  from an upper row, $C_{j}$ and $C_{k}$, multiplied by
  corresponding phase change.}
\label{fig:Fig3}
\end{figure}

\begin{figure}
\epsfig{file=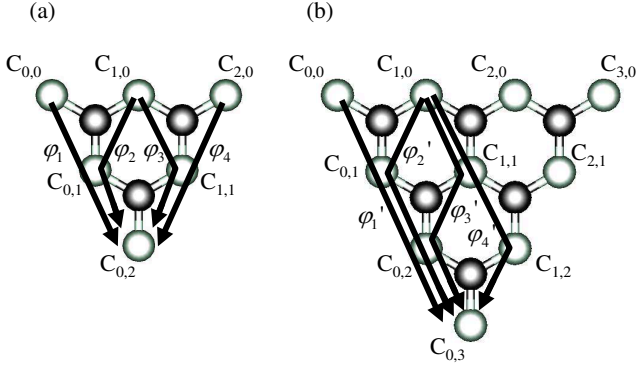,width=3.4in}
\caption{(Color online) (a) A hexagonal benzene ring with three
  auxiliary corner atoms added. Each $A$-type site is described by two
  numbers $\{n,m\}$. Black arrows indicate phase changes related to
  the paths going from an upper row of atoms, with indices $\{n,0\}$,
  to an atom from in the bottom, $C_{0,2}$. (b) Triangular zigzag
  graphene quantum dot with $N_{ed}=2$ atoms on one edge. There is
  one path going from site $C_{0,0}$ to $C_{0,3}$, and three paths
  going from site $C_{1,0}$ to $C_{0,3}$. The number of paths
  connecting a site $\{n,m\}$ with a site from the top $\{n+j,0\}$ can
  be described by binomial coefficient $N_{path}(n,m,n+j)={m \choose
    j}$, $0\leq j\leq m$.}
\label{fig:Fig4}
\end{figure}

According to Eq. \ref{singul}, the
coefficients $C_{j}$ corresponding to one type orbitals localized around the
second type site $i$ obey
\begin{eqnarray}
t\sum_{\left\langle i,j\right\rangle}C_{j}e^{i\varphiup_{ij}}=0,
\label{cond}
\end{eqnarray}
where the summation is over $j$-th nearest neighbors of an atom $i$. In other
words, the sum of coefficients multiplied by a phase $e^{i\varphiup_{ji}}$
gained by going from one type site $i$ to the other type site $j$ around each
site $i$ must vanish. For the $i$-th $B$-type site plotted on the left in Fig. \ref{fig:Fig3}, Eq. (\ref{cond}) gives
\begin{eqnarray}
C_{j}e^{i\varphiup_{ij}}+C_{k}e^{i\varphiup_{ik}}+C_{l}e^{i\varphiup_{il}}=0,
\label{cond2}
\end{eqnarray}
where phases $\varphiup_{ij},\varphiup_{ik},\varphiup_{il}$ are given by Eq. (\ref{phaseij}). Using a fact that $\varphiup_{ik}=-\varphiup_{ki}$ for arbitrary $i$ and $k$, Eq. (\ref{cond2}) can be written as
\begin{eqnarray}
C_{l}=-\left(C_{j}e^{-i\varphiup_{jl}}+C_{k}e^{-i\varphiup_{kl}}\right),
\label{cond4}
\end{eqnarray}
where $\varphiup_{jl}=\varphiup_{ji}+\varphiup_{il}$ and
$\varphiup_{kl}=\varphiup_{ki}+\varphiup_{il}$ correspond to phase changes
going from $A$-type sites $k$ to $j$, and $l$ to $j$, respectively, through
$B$-type site $i$ (see right part in Fig. \ref{fig:Fig3}). Thus, in analogy with
the zero magnetic field case \cite{Potasz+10}, a coefficient from a given row
can be expressed as a sum of two coefficients from an upper lying row,
$C_{j}$ and $C_{k}$ on the right in Fig. 3. The effect of the
magnetic field is incorporated in the extra phase gained by going from a given
site from an upper row of atoms to a lower one. For a reason which will become
 clear later, instead of using indices $i$, each $A$-type site will be
labeled by two integer numbers, ${i}=\{n,m\}$. The first index, $n$,
corresponds to an atom  number  in a given row counted from left to
right, and the second one, $m$, corresponds to the row number.
Let us illustrate
our methodology  on a hexagonal benzene ring  with three auxiliary $A$-type  atoms with indices $C_{0,0}$, $C_{2,0}$ and
$C_{0,2}$, shown in Fig. \ref{fig:Fig4}(a). Eq. (\ref{cond4}) can be used to
obtain coefficients $C_{0,1}$ from $C_{0,0}$ and $C_{1,0}$, and $C_{1,1}$ from
$C_{1,0}$ and $C_{2,0}$. Next, using $C_{0,1}$ and $C_{1,0}$, one obtains
coefficient $C_{0,2}$,
\begin{eqnarray}
C_{0,2}=C_{0,0}e^{-i\varphiup_{1}}+C_{1,0}\left(e^{-i\varphiup_{2}}+e^{-i\varphiup_{3}}\right)
+C_{2,0}e^{-i\varphiup_{4}},
\label{eq5}
\end{eqnarray}
with phase changes $\varphiup_{i}$, $i=1,2,3,4$, shown as black arrows in
Fig. \ref{fig:Fig4}(a). The paths related to phase changes $\varphiup_{i}$ go
through intermediate atomic sites, e.g.,  for $\varphiup_{1}$ the path goes from
a site $C_{0,0}$ to $C_{0,1}$ through an intermediate $B$-type atomic site, and
next from a site $C_{0,1}$ to $C_{0,2}$ through connecting  $B$-type
atomic site.
According to Eq. (\ref{eq5}) and Fig. \ref{fig:Fig4}(a), there is
one path connecting $C_{0,0}$ and $C_{0,2}$, one connecting $C_{2,0}$ and
$C_{0,2}$, but there are two paths around a hexagonal benzene ring connecting
coefficients $C_{1,0}$ and $C_{0,2}$. We have shown that the coefficient in
the bottom, $C_{0,2}$, can be expressed as a linear combination of
coefficients from the top row, $C_{n,0}$. We will now  demonstrate that all
coefficient in arbitrary size triangles can be expressed
in terms of coefficients $C_{n,0}$.

In Fig. \ref{fig:Fig4}(b) a small triangle with $N_{ed}=2$ atoms on the one
edge is plotted. Three auxiliary atoms with coefficients $C_{0,0}$, $C_{3,0}$,
and $C_{0,3}$ were added.
The total number of atoms is $N=16$. In a similar way to the procedure used to
obtain Eq. (\ref{eq5}), a coefficient $C_{0,3}$ can be expressed as a sum of
coefficients from the top. Here, from coefficients $C_{0,0}$ and $C_{3,0}$ to
$C_{0,3}$ there is only one path for each coefficient, and three paths for
each coefficient connecting $C_{1,0}$ to $C_{0,3}$, and $C_{2,0}$ to
$C_{0,3}$. For transparency, only for the first two coefficients from the
left ($C_{0,0}$ and $C_{1,0}$) paths are plotted in
Fig. \ref{fig:Fig4}(b). The number of paths from a given site in the upper row
of atoms to lower lying atomic sites corresponds to numbers from a Pascal
triangle, $\{1,2,1\}$ for coefficient $C_{0,2}$, shown in
Fig. \ref{fig:Fig4}(a), and $\{1,3,3,1\}$ for coefficient $C_{0,3}$, shown
for the first two coefficients from the left in Fig. \ref{fig:Fig4}(b). The number
of paths connecting a site $\{n,m\}$ with a site from the top $\{n+j,0\}$ can
be described by binomial coefficient $N_{path}(n,m,n+j)={m \choose j}$, $0\leq
j\leq m$. The general form for an arbitrary coefficient expressed in
coefficients from the top row can be written as
\begin{eqnarray}
C_{n,m}=(-)^{m}\sum_{j=0}^{m}\sum_{i=1}^{m \choose j}C_{n+j,0}e^{-i\varphiup_{n+j}(i)},
\label{eq6}
\end{eqnarray}
where two numbers $n$ and $m$ satisfy condition $0<n,m<N_{ed}+1$, and
$\varphiup_{n+j}(i)$ is a path-dependent phase change from a site $\{n+j,0\}$
to $\{n,m\}$. One can note that in the absence of a magnetic field
$\varphiup_{n+j}(i)=0$ and Eq. (\ref{eq6}) reduces to Eq.(2) from
Ref.\cite{Potasz+10}.

The summation over all possible paths in Eq. (\ref{eq6}) is not practical. We
now show a way of reducing the number of paths to only one. We use the fact that
a phase change corresponding to a closed path around a hexagon is by
definition $\varphiup_{c}=2\pi\phi/\phi_0$. The sum of two exponential terms
standing next to coefficient $C_{1,0}$ in Eq. (\ref{eq5}) can be written as
\begin{eqnarray}
\label{expon1}
e^{-i\varphiup_{2}}+e^{-i\varphiup_{3}}=\left(e^{i(\varphiup_{3}-\varphiup_{2})}+1\right)e^{-i\varphiup_{3}}
=\left(e^{2\pi i\frac{\phi}{\phi_0}}+1\right)e^{-i\varphiup_{3}},
\end{eqnarray}
where $\varphiup_{3}-\varphiup_{2}=2\pi\phi/\phi_0$ is a closed path around a
single hexagon, see Fig. \ref{fig:Fig4}(a). Similarly for three exponential
terms corresponding to paths connecting $C_{10}$ and $C_{03}$, shown in
Fig. \ref{fig:Fig4}(b), one can write
\begin{eqnarray}
\nonumber
e^{-i\varphiup_{2}'}+e^{-i\varphiup_{3}'}+e^{-i\varphiup_{4}'}&=&
\left(e^{i(\varphiup_{4'}-\varphiup_{2'})}+e^{i(\varphiup_{4'}-\varphiup_{3'})}+1\right)e^{-i\varphiup_{4'}}\\
&=&\left(e^{2\pi i(2\frac{\phi}{\phi_0})}+e^{2\pi i\frac{\phi}{\phi_0}}+1\right)e^{-i\varphiup_{4'}},
\label{expon2}
\end{eqnarray}
where $(\varphiup_{4'}-\varphiup_{2'})$ circles two hexagons and
$(\varphiup_{4'}-\varphiup_{3'})$ only one, see Fig. \ref{fig:Fig4}(b). Note
that phases $\varphiup_{3}$ in Eq. (\ref{expon1}) and $\varphiup_{4'}$ in
Eq. (\ref{expon2}) correspond to paths going  on the right edge of the triangle. The sum
of exponential terms of type $e^{2\pi i(j\frac{\phi}{\phi_0})}$ with
$j$-integer in Eq. (\ref{expon1}) and Eq. (\ref{expon2}) forms geometric
series which can be written as
\begin{eqnarray}
\sum_{j=0}^{k}e^{2\pi i(j\frac{\phi}{\phi_0})}=\frac{1-e^{2\pi i(k+1)\frac{\phi}{\phi_0}}}{1-e^{2\pi i\frac{\phi}{\phi_0}}},
\label{expon3}
\end{eqnarray}
with $k$ determined by the number of encircled benzene rings, and $k+1={m
  \choose j}$ is a number of paths connecting site $\{n+j,0\}$ to $\{n,m\}$,
$k=1$ in Eq. (\ref{expon1}) and $k=2$ in Eq. (\ref{expon2}), see
Fig. \ref{fig:Fig4}. Using Eq. (\ref{expon3}), the number of paths in
Eq. (\ref{eq6}) can be reduced to only one. Eq. (\ref{eq6}) can be written as
\begin{eqnarray}
C_{n,m}=(-)^{m}\sum_{j=0}^{m}C_{n+j,0}\frac{1-e^{2\pi i{m \choose j}\frac{\phi}{\phi_0}}}{1-e^{2\pi i\frac{\phi}{\phi_0}}}e^{-i\varphiup_{n+j}},
\label{eq7}
\end{eqnarray}
where $\varphiup_{n+j}$ is the phase corresponding to the path on the
right edge connecting site $\{n+j,0\}$ and $\{n,m\}$. The coefficients
$C_{n,m}$ for all $A$-type atoms in the triangle are expressed as a
linear combination of coefficients corresponding to atoms on one edge,
i.e., $C_{j,0}$. There are $N_{ed}+2$ coefficients in an upper row of
atoms, $C_{j,0}$, with $0<j<N_{ed}+1$, which gives $N_{ed}+2$
independent solutions. Applying three boundary conditions
corresponding to auxiliary atoms,
$C_{0,0}=C_{N_{ed}+1,0}=C_{0,N_{ed}+1}=0$, leaves only $N_{ed}-1$
solutions, which corresponds to the number of zero-energy states,
similar to the result obtained in the absence of a magnetic field in
Ref. \onlinecite{Potasz+10}. We note that the solutions given by
Eq. (\ref{eq7}) are smooth functions of magnetic field, and exist for
any value of $\phi$. Thus they do not include zero-energy solutions
corresponding to the crossing of conduction and valence states with
$E=0$, e. g., for $\phi/\phi_0\simeq 0.11$ for the triangular dot with
$N_{ed}=8$ and $N=97$ atoms, see Fig. \ref{fig:Fig1}. We investigate
this issue by analyzing $B$-type atoms.

\subsection{Prediction of crossings of valence and conduction states with $E=0$}

\begin{figure}
\epsfig{file=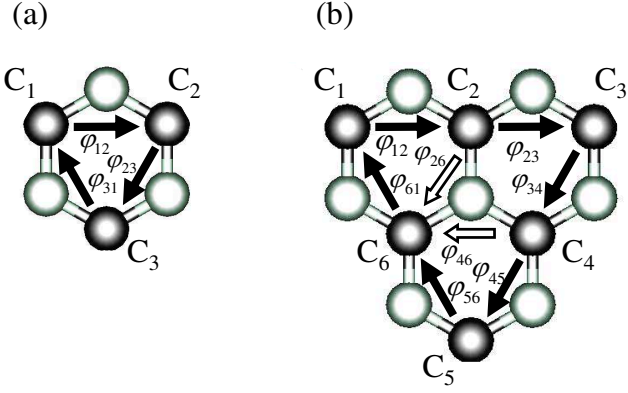,width=3.4in}
\caption{(Color online) (a) A hexagonal benzene ring with coefficients
  $C_{i}$ assigned to each $B$-type site. Black arrows indicate phase
  changes related to the paths going from one $B$-type site to
  another. (b) Triangular zigzag graphene quantum dot with $N_{ed}=2$
  atoms on the one edge. Black arrows indicate phase changes related
  to the paths going from one $B$-type site to another along
  triangular three edges. White arrows indicate phase changes related
  to the paths going through the center.}
\label{fig:Fig5}
\end{figure}

\begin{figure}
\epsfig{file=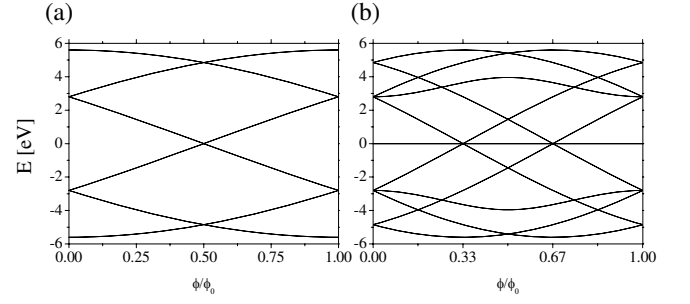,width=3.4in}
\caption{Energy spectrum as function of magnetic flux for (a) a
  hexagonal benzene ring and (b) a triangular graphene quantum dot
  with $N_{ed}=2$ atoms on the one edge and a total of $N=13$ atoms. A
  crossing of conduction and valence states with $E=0$ occurs for (a)
  $\phi/\phi_0=1/2$ and (b) $\phi/\phi_0=1/3$ and $\phi/\phi_0=2/3$.}
\label{fig:Fig6}
\end{figure}

We consider the solution of Eq. (\ref{singul}) corresponding to
wavefunction localized only on $B$-type atoms. In Fig. \ref{fig:Fig5}
the same structures as in Fig. \ref{fig:Fig4} without auxiliary corner
atoms are shown with coefficients assigned to $B$-type atoms. For
simplicity, only one index for each coefficient is used.
According to Eq. \ref{cond}, for a benzene ring plotted in Fig. \ref{fig:Fig5}(a) we can write
\begin{eqnarray}
\label{ap_eq1a}
C_{2}=-C_{1}e^{-i\varphiup_{12}},\\
\label{ap_eq1b}
C_{3}=-C_{2}e^{-i\varphiup_{23}},\\
C_{1}=-C_{3}e^{-i\varphiup_{31}},
\label{ap_eq1c}
\end{eqnarray}
where phase changes from site $i$ to $j$, $\varphiup_{ij}$, are indicated in
Fig. \ref{fig:Fig5}(a). Eq. (\ref{ap_eq1a}) can be substituted into
Eq. (\ref{ap_eq1b}), and next Eq. (\ref{ap_eq1b}) into Eq. (\ref{ap_eq1c}),
giving
\begin{eqnarray}
C_{1}=C_{1}(-1)^3e^{-i(\varphiup_{12}+\varphiup_{23}+\varphiup_{31})},
\label{ap_eq2}
\end{eqnarray}
which is satisfied for arbitrary $C_{1}$. Eq. \ref{ap_eq2} leads to a following condition
\begin{eqnarray}
\varphiup_{12}+\varphiup_{23}+\varphiup_{31}+\pi=2\pi k,
\label{ap_eq4}
\end{eqnarray}
with $k=0,\pm 1,\pm 2,...$. A phase change in Eq. (\ref{ap_eq4}) corresponds
to a closed path around a single hexagon,
$\varphiup_{12}+\varphiup_{23}+\varphiup_{31}=2\pi\phi/\phi_0$. A condition
for crossing of a valence and conduction states with $E=0$ is
\begin{eqnarray}
\phi/\phi_0=k-1/2.
\label{ap_eq7}
\end{eqnarray}
In order to confirm validity of Eq. (\ref{ap_eq7}), we show the energy
spectrum of a benzene ring as a function of a magnetic field in
Fig. \ref{fig:Fig6}(a). The crossing of energy levels at $E=0$ occurs
for $\phi/\phi_0=1/2$, in agreement with Eq. (\ref{ap_eq7}).

We carry out a similar derivation for triangular zigzag graphene
quantum dot with $N=13$ carbon atoms and $N_{ed}=2$ atoms on the one
edge, shown in Fig. \ref{fig:Fig5}(b).  A coefficient from the left
upper corner, $C_1$, determines a coefficient $C_2$,
\begin{eqnarray}
\label{ap_eq8}
C_{2}=-C_{1}e^{-i\varphiup_{12}}.
\end{eqnarray}
Next, a coefficient $C_3$ can be determined by a coefficient $C_2$
\begin{eqnarray}
\label{ap_eq9}
C_{3}=-C_{2}e^{-i\varphiup_{23}}
\end{eqnarray}
and combining with Eq. (\ref{ap_eq8}) gives
\begin{eqnarray}
\label{ap_eq10}
C_{3}=(-1)^2C_{1}e^{-i(\varphiup_{12}+\varphiup_{23})}=(-1)^2C_{1}e^{-i\varphiup_{13}} .
\end{eqnarray}
Going in this way along the three edges of the triangle a closed loop,
shown with black arrows in Fig. \ref{fig:Fig5}(b), can be created. In
the
case of $N_{ed}=2$ shown in Fig. \ref{fig:Fig5}(b), one goes through all $B$-type
coefficients, while in larger triangles one goes only through outer
coefficients. Thus, all outer $B$-type coefficients can be expressed by one chosen
coefficient, $C_1$ in this case. The loop from Fig. \ref{fig:Fig5}(b) can be
written
\begin{eqnarray}
\label{ap_eq11}
C_{1}=(-1)^6C_{1}e^{-i(\varphiup_{16}+\varphiup_{61})}=C_{1}e^{6\pi i-2\pi i(3\phi/\phi_0)},
\end{eqnarray}
where the phase change $\varphiup_{16}=\sum_i^5\varphiup_{i,i+1}$, and we used
a fact that the total phase change corresponds to a closed loop around three
benzene rings, $\varphiup_{16}+\varphiup_{61}=2\pi
(3\phi/\phi_0)$. Eq. (\ref{ap_eq11}) gives a condition
\begin{eqnarray}
\label{ap_eq14}
2k\pi=6\pi-6\pi\phi/\phi_0,
\end{eqnarray}
and finally
\begin{eqnarray}
\label{ap_eq15}
\phi/\phi_0=\frac{3-k}{3}.
\end{eqnarray}
with $k=0,\pm 1,\pm 2,...$. Eq. (\ref{ap_eq15}) can be extended to different
size triangles. The number of benzene rings in a triangle is
$N_b=N_{ed}(N_{ed}+1)/2$, and Eq. (\ref{ap_eq15}) can be written as
\begin{eqnarray}
\label{ap_eq15n}
\phi/\phi_0=\frac{3N_{ed}-2k}{N_{ed}\left(N_{ed}+1\right)}.
\end{eqnarray}
For the triangle with $N_{ed}=2$, Eq. (\ref{ap_eq15}) predicts crossings for
$\phi/\phi_0=0,\pm 1/3, \pm 2/3, 1,..$ but according to Fig. \ref{fig:Fig6}(b)
there are no crossings for $\phi/\phi_0=0$ and $\phi/\phi_0=1$. This is
related to an extra condition in the center of the triangle, for coefficients
$C_2$, $C_4$, and $C_6$. Phase changes between these coefficients are
indicated by white arrows in Fig. \ref{fig:Fig5}(b). We can write
\begin{eqnarray}
\label{ap_eq16}
C_{6}=-\left(C_{2}e^{-i\varphiup_{26}}+C_{4}e^{-i\varphiup_{46}}\right)
\end{eqnarray}
and also
\begin{eqnarray}
\label{ap_eq17}
C_{6}=-C_{1}e^{-i(-\varphiup_{61})}=-C_{1}e^{i\varphiup_{61}},\\
\nonumber
C_{2}=-C_{1}e^{-i\varphiup_{12}},\\
\nonumber
C_{4}=(-1)^3C_{1}e^{-i\varphiup_{14}},
\end{eqnarray}
where the phase change $\varphiup_{14}=\sum_i^3\varphiup_{i,i+1}$. Combining Eq. (\ref{ap_eq16}) and Eq. (\ref{ap_eq17}) we get
\begin{eqnarray}
\label{ap_eq18}
-C_{1}e^{i\varphiup_{61}}=-\left(-C_{1}e^{-i(\varphiup_{12}+\varphiup_{26})}
+(-1)^3C_{1}e^{-i(\varphiup_{14}+\varphiup_{46})}\right),
\end{eqnarray}
which gives
\begin{eqnarray}
\label{ap_eq19}
-1=e^{-i(\varphiup_{12}+\varphiup_{26}+\varphiup_{61})}
+e^{-i(\varphiup_{14}+\varphiup_{46}+\varphiup_{61})}.
\end{eqnarray}
With help of Fig. \ref{fig:Fig5}(b), we can notice $\varphiup_{12}+\varphiup_{26}+\varphiup_{61}=2\pi\phi/\phi_0$ and $\varphiup_{14}+\varphiup_{46}+\varphiup_{61}=2\pi(2\phi/\phi_0)$. Thus, we can write
\begin{eqnarray}
\label{ap_eq21}
1+e^{-2\pi i\phi/\phi_0}+e^{-2\pi i(2\phi/\phi_0)}=0
\end{eqnarray}
or using a sum of geometric series
\begin{eqnarray}
\label{ap_eq22}
\frac{1-e^{-2\pi i(3\phi/\phi_0)}}{1-e^{-2\pi i\phi/\phi_0}}=0.
\end{eqnarray}
Eq. (\ref{ap_eq22}) gives a solution for $-2\pi (3\phi/\phi_0)=2\pi k$,
$k$-integer, and finally $\phi/\phi_0=-k/3$, but with an extra condition
$\phi/\phi_0\neq l$, with $l=0,\pm 1, \pm 2,..$ due to a denominator. This is
in agreement with Fig. \ref{fig:Fig6}(b). We note that for all triangles, the
prediction of crossings of conduction and valence states with $E=0$ given by
Eq. (\ref{ap_eq15n}) has to be supported by extra conditions from equations
for coefficients from the center of the triangle. For example, for the
triangle with $N_{ed}=8$, the first crossing occurs for $\phi/\phi_0=1/9$,
while incomplete condition given by Eq. (\ref{ap_eq15n}) predicts the first
crossing for $\phi/\phi_0=1/36$, and the fourth crossing for
$\phi/\phi_0=1/9$.
\begin{figure}
  \epsfig{file=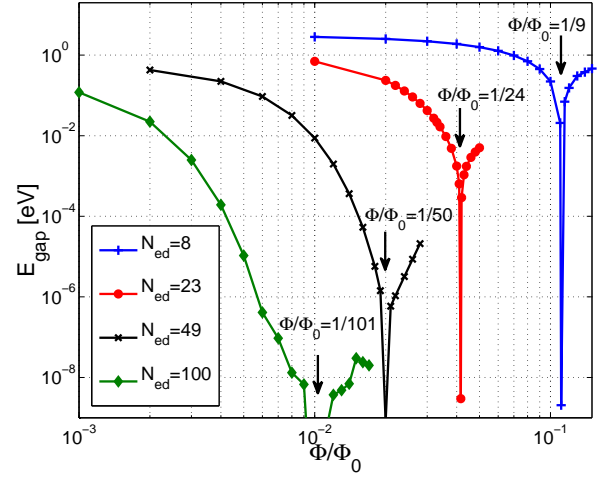,width=3.2in}
\caption{Energy spectrum as a function of magnetic flux for 
different sizes of triangular zigzag quantum dots, showing that first zero
energy crossing occurs at $\phi/\phi_0=1/(N_{ed}+1)$.
}
\label{fig:Fig7}
\end{figure}

An interesting prediction of Eq.(\ref{ap_eq15}) is that the zero
energy crossing values of $\phi/\phi_0$ should scale as $\sim 1/N_{ed}$ for
large $N_{ed}$. In order to check numerically the size dependence of the
position of the first crossing, in Fig. \ref{fig:Fig7} we show the
energy gap as a function of $\phi/\phi_0$ for different $N_{ed}$
obtained by diagonalization of the tight-binding
Hamiltonian. Strikingly,  we find that the first crossing always
occurs at $\phi/\phi_0=1/(N_{ed}+1)$ for all the values of $N_{ed}$
that we have looked at. This is consistent with Eq.(\ref{ap_eq15})
with $k=N_{ed}$. Extrapolating this result to larger structures, it
would take a magnetic field value of $\sim 10$ Tesla for a quantum dot with
$N_{ed}=4000$ to reach the first zero energy crossing.

\begin{figure}
  \epsfig{file=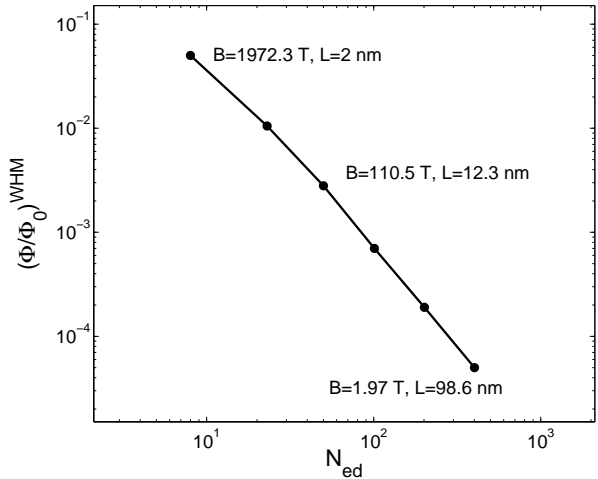,width=3.2in}
\caption{Width at half maximum of the flux dependence of the energy gap
as a function of $N_{ed}$, showing a $\sim N_{ed}^{-2}$ dependence at large
$N_{ed}$. For few points, corresponding magnetic field values and linear sizes
are given in Tesla and nm respectively.
}
\label{fig:Fig8}
\end{figure}

However, for large quantum dots ($N_{ed}>100$, or linear size $L>25$ nm) it
becomes increasingly  difficult to pinpoint numerically the position
of the zero energy crossing due to smallness of the energy gap around
the crossing and numerical accuracy. Another quantity of interest is
the width at half maximum (WHM) of the flux dependence of the energy
gap. In Fig.  \ref{fig:Fig8} we plot the WHM as a function of
$N_{ed}$. Unlike the first  crossing point which scales as
$N_{ed}^{-1}$, the WHM scales as $\sim N_{ed}^{-2}$ for large $N_{ed}$,
thus much faster. In Fig. \ref{fig:Fig8} the largest structure that we
looked at has $N=161601$ atoms ($N_{ed}=401$, $L=98.6$ nm) for which the WHM occurs
at a magnetic field value of $B=1.97$ Tesla.

\section{SHAPE AND EDGE DEPENDENCE OF THE ENERGY GAP IN A MAGNETIC FIELD}

\begin{figure}
\epsfig{file=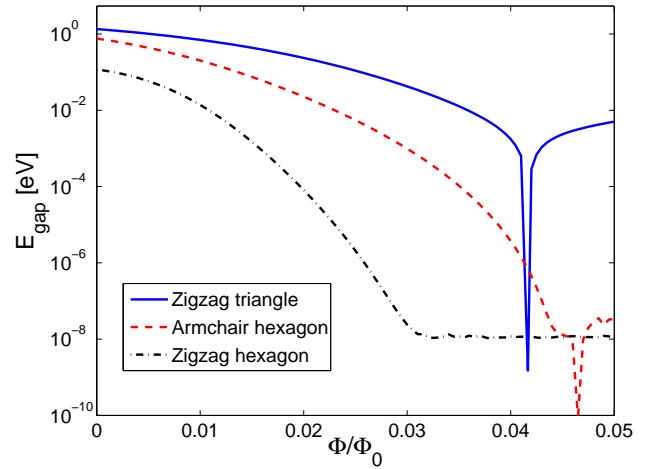,width=3.4in}
\caption{The energy gap as function of the magnetic
flux for triangular zigzag quantum dot with $N=622$ atoms (solid line), 
hexagonal armchair quantum dot with $N=546$ atoms (dashed line), 
hexagonal zigzag quantum dot with $N=600$ atoms (dot-dashed line).}
\label{fig:Fig9}
\end{figure}

We discussed above the magnetic field closing of the energy gap in
triangular graphene quantum dots.  In Fig. \ref{fig:Fig9}, we analyze
the evolution of the energy gap in graphene quantum dots with
different shapes and edges in a perpendicular magnetic field. The
energy gaps as a function of a magnetic field obtained by
diagonalizing  Hamiltonian given by Eq. (\ref{HTB}) are shown for
three different types of quantum dots; zigzag triangle, zigzag
hexagon, and armchair hexagon. All three structures have  similar
sizes, consisting of $N\simeq 600$ atoms with area $S\simeq 14$
nm$^2$.  The energy gap corresponds to the difference between the
energy of the lowest state from the empty conduction states and the
highest state from the doubly occupied valence states. In the absence
of magnetic field, the zigzag triangular graphene quantum dot has a
significantly larger gap then for hexagonal armchair and zigzag dots
as discussed in Ref.\onlinecite{Guclu+10}. The functional form of the
gap closure of different types of structures has significant differences
as well, as seen in  Fig. \ref{fig:Fig9}.
When the magnetic field increases, the energy gap closes for all
structures. Although the hexagonal zigzag structure has slightly
smaller size, the gap decays fastest showing a different behavior
than the $\sim N_{ed}^{-2}$ scaling shown earlier for the triangular
zigzag structure. Moreover, after reaching a plateau close to zero
($\sim 10^{-8}$) the hexagonal zigzag quantum dot shows no more
structures, {\it i.e.} no zero energy crossings, unlike the two other
quantum dots. We note that for the hexagonal zigzag  structure the gap
comes from closure of the edge-like states (which have finite energies
unlike the triangular zigzag structure). This shows that the zero
crossings are characteristics of bulk-like states. \\

\section{CONCLUSIONS}
The  electronic properties of triangular graphene quantum dots with
zigzag edges and broken sublattice symmetry in the presence of
perpendicular external magnetic field were described.  It was shown
that  the degenerate shell of zero-energy states in the middle of the
energy gap is immune to the magnetic field in analogy to the $n=0$
Landau level of bulk graphene.  An analytical solution for zero-energy
states in the magnetic field was derived. The energy gap was shown to
close with increasing magnetic field, reaching zero at special values
of the magnetic field. The gap closing was found independent of
quantum dot size, shape, and edge termination.

\section*{ACKNOWLEDGMENT}
The authors thank  NSERC, the Canadian Institute for Advanced Research
and T\"UBITAK for support. P.P thanks for fellowship within "Mistrz"
program from The Foundation for Polish Science.


\end{document}